\documentclass[aps,prl,twocolumn,groupedaddress]{revtex4-1}
\usepackage[utf8]{inputenc}
\usepackage[english]{babel}
\usepackage{amsmath,graphicx,enumerate}

\begin{document}

%\preprint{}

\title{Comment on ``Entropy Production and Fluctuation Theorems for Active Matter''} 


\author{Lorenzo Caprini}
\affiliation{Gran Sasso Science Institute (GSSI), Via. F. Crispi 7, 67100 L’Aquila, Italy}
\author{Umberto Marini Bettolo Marconi}
\affiliation{Scuola di Scienze e Tecnologie, Università di Camerino - via Madonna delle Carceri, 62032, Camerino, Italy}
\author{Andrea Puglisi}
\affiliation{Istituto dei Sistemi Complessi - CNR and Dipartimento di Fisica, Universit\`a di Roma Sapienza, P.le Aldo Moro 2, 00185, Rome, Italy}
\author{Angelo Vulpiani}
\affiliation{Dipartimento di Fisica, Università Sapienza - p.le A. Moro 2, 00185, Roma, Italy}
\affiliation{Centro Interdisciplinare "B.Segre", Accademia dei Lincei, Roma, Italy}
%\thanks{}


% \date{\today}

% \pacs{}
%\keywords{}

\maketitle

% the lengths of Comments and Replies are restricted to 750 words each (roughly one journal page).
% https://journals.aps.org/prl/authors/comments-physical-review-letters

In~\cite{mandal}, Mandal, Klymko and DeWeese challenge the existing
formula for entropy production for a renowned model (AOUP in their
Letter) of active particles~\cite{fodor,maggi}. In this Comment, we
question the central results of~\cite{mandal}.

First, we show that the main result in~\cite{mandal}, that is their
Eq.~(9), is not correct. In order to prove this, one may directly
analyze Eq. (2a) of~\cite{mandal}, without the need of any
transformation of variables. This circumvents the indeterminacy of
the time-reversal operation for the Gaussian random force, ${\bf v}_i$,
which is equal to the sum of two quantities having opposite
time-reversal parities, as Eq. (2a) declares. The path probabilities
induced by the colored noise are calculated without any ambiguity,
see~\cite{zamponi} or~\cite{crisanti}. Such a
calculation,  in the case of a single particle in one
dimension, $\Gamma(t) \equiv x(t)$, returns
\begin{equation}
P[\Gamma] \propto \exp\left[-\frac{1}{2}\int dt \int ds \; v(t) T^{-1}(t-s) v(s)\right]
\end{equation}
where we have defined
$T^{-1}(t)=\frac{1}{2D}\delta(t)\left(1-\tau^2\frac{d^2}{dt^2}\right)$
in such a way that $\int ds' T^{-1}(t-s') \langle v(s')v(s)
\rangle=\delta(t-s)$. With algebra one gets the following
formula for the entropy production $\Sigma[\Gamma] = \log
\frac{P[\Gamma]}{P[\Gamma^r]}$, i.e. the only possible prescription
for the AOUP system (and in fact Eq. (7) of~\cite{mandal} has not an
equivalent in the overdamped equation):
\begin{equation}
\Sigma[\Gamma] = - \mu\int_0^t \;ds \dot{x}(s) (T^{-1}\ast \Phi')(s) + \Phi'(s)(T^{-1}\ast \dot{x})(s), 
\end{equation}
where $*$ stands for the convolution operation. Performing algebraic manipulations one finally gets:
\begin{equation} \label{zamp}
\Sigma[\Gamma] = b.t. + \frac{\mu\tau^2}{2D}\int^t  \dot{x}^3(s) \Phi'''(s) ds,
\end{equation}
where b.t. denotes boundary terms. This result is clearly
different from Eq. (9) of~\cite{mandal}. Neglecting the b.t.,
Eq.~\eqref{zamp} coincides with the results in [2, 3], which have been
obtained through the underdamped mapping (analogous to Eqs. (3) of
[1]) and adopting a different time-reversal operation for the
non-equilibrium force. Remarkably, Eq.~\eqref{zamp} does not require
any prescription of such kind: for this reason it seems to us
indisputable.

Second, we contest the identification of $-p/(\mu m) + \sqrt{2/(\mu
  \beta)}\eta$ with a thermal bath, which follows from a crucial
confusion between $p=m \dot{x}$ and real particles’ momentum. Based on
this, the authors of~\cite{mandal} state that the total energy is
$E=\frac{p^2}{2m}+\Phi(x)$. However, the AOUP system is different: it
is described by an overdamped equation where the real particles' mass
and momentum are unknown and their kinetic energy, in general, is not
$\frac{p^2}{2m}$.  The heat, the choice of the time-reversal and the
whole ``derivation'' of Eq. (7)~\footnote{Section III of the
  Supplemental Material of~\cite{mandal}}, all stem from such a wrong
identification of mass, momenta and kinetic energy.  A consequence is
seen when the potential $\Phi$ is removed, e.g. by considering a
single particle and no external forces. In this case the average heat
exchange, Eq. (5) of~\cite{mandal} and entropy production, Eq. (9)
of~\cite{mandal}, both vanish even when $\tau > 0$: the model results
at equilibrium even if it describes an active particle.

Third, we observe that a central application of their main result, the
detailed fluctuation relation, Eq.~(12) in~\cite{mandal}, cannot be
verified in experiments, since it involves a measurement of
$P^r(\Sigma^r)$, the probability of entropy production according to a
different dynamics, Eq. (7), which is not known to represent any
realizable system.

After having shown how to remove certain ambiguities in the AOUP
model, we sketch their origin~\cite{maggi,entropy}. The AOUP model
only represents a coarse-grained level of description, where the
variables $\{\dot{x}_i\}$ are not the real velocities of the particles
and the fluctuating force (white noise) of the {\em real} thermal bath
has been neglected.  For this reason there is no way to take into
account the energetic and entropic exchanges with the physical
thermostat and it is not surprising to observe zero entropy production
in some special cases, even if physically one expects it to be
positive: that is a possible outcome of
coarse-graining~\cite{villamaina,crisanti,cerino} which occurs also in
[1] when $\Phi=0$. This is however not a reason to identify the
non-conservative self-propulsion force, or part of it, as a thermal
bath force, a choice which is physically wrong and leads - as we
showed - to inconsistent results. A derivation of an entropy production
formula without resorting to any arbitrary prescription for
time-reversal, as sketched in Eq.~\eqref{zamp} above, is the simplest
way to settle the dispute.

\end{document}